\documentstyle[sprocl,epsfig]{article}

\def\Journal#1#2#3#4{{#1} {\bf #2}, #3 (#4)}

\def\PRL{\em Phys. Rev. Lett.}
\def\PRD{{\em Phys. Rev.} D}

\def\PHZ{{\em Physik} Z.}
\def\AJ{\em AJ}
\def\APJ{\em ApJ}
\def\MN{\em MNRAS}
\def\PASJ{\em PASJ}
\def\AA{\em A\&A}

\def\be{\begin{equation}}
\def\ee{\end{equation}}
\def\bea{\begin{eqnarray}}
\def\eea{\end{eqnarray}}

\begin{document}

\title{ON GRAVITOMAGNETIC PRECESSION AND QPO IN BLACK HOLE CANDIDATES}
\author{ WEI CUI }
\address{Center for Space Research, Massachusetts Institute of Technology,\\
Cambridge, MA 02139, USA\\E-mail: cui@space.mit.edu}
\author{ WAN CHEN \footnote{also Department of Astronomy, University of Maryland, College Park, MD 20742, USA}}
\address{ NASA/Goddard Space Flight Center, Code 661, \\
Greenbelt, MD 20771, USA\\E-mail: chen@milkyway.gsfc.nasa.gov}
\author{ S.N. ZHANG }
\address{Department of Physics, University of Alabama-Huntsville,\\
Huntsville, AL 35812, USA\\E-mail: Shuang.Zhang@msfc.nasa.gov}

\maketitle\abstracts{The X-ray observation of black hole candidates
provides a valuable tool to probe regions very close to the central  
black hole, where strong-field relativistic effects become important.
Recent studies have shown that these effects seem to manifest 
themselves in the observed X-ray spectrum, in terms of the shape of 
X-ray continuum and the profile of emission lines, and in the X-ray 
light curves, in terms of certain quasi-periodic oscillations (QPOs). 
The latter will be the focus of this paper. We will review the proposed
observational evidence for gravitomagnetic precession in black hole 
candidates, in light of new observational data and alternative models.
Quantitative comparison will be made between the data and the models. 
We will comment on recent theoretical and observational efforts to 
address whether the gravitomagnetic precession of accreted, orbiting 
matter around a rotating black hole is a viable process for producing 
the observed QPOs. The results are inconclusive. We will mention a few 
areas where further progress can be made to possibly shed more light 
on the issue.}

\section{Background}
\subsection{X-ray Binaries}
X-ray binaries are among the brightest X-ray sources in the sky. An 
X-ray binary consists of a compact object (neutron star or black hole) 
and a normal companion star, orbiting around each other. At certain 
stage of the binary evolution, the separation of the two stars becomes 
so close that material from the upper atmosphere of the companion star 
begins to flow toward the compact object under the influence of the 
latter's gravity. This process is known
as mass accretion due to ``Roche-lobe overflow''. The accreted 
matter, carrying large amount of angular momentum from the orbital
motion, circulates around the compact object and forms an accretion 
disk. As the matter spirals in toward the compact object, due to 
angular momentum losses from viscous processes, its gravitational 
energy is converted into heat. Depending upon the mass accretion rate,
the temperature of the inner accretion disk can reach more than one 
million degrees so that X-rays are produced. Therefore, the X-ray 
observation of such 
sources provides a valuable tool to probe regions very close 
to the compact object, where relativistic effects are strong thus 
important. It should be noted, however, that Roche-lobe overflow 
does not always occur. For some sources (especially 
those with a massive companion star), the accretion process may simply 
involve the capture of stellar wind from the companion star by the 
compact object. A small accretion disk can still form in such ``wind-fed'' 
systems, due to any residual angular momentum of the captured matter. 

\subsection{Black Hole Binaries}
In some cases, the companion star is visible optically. By carefully
monitoring its orbital motion, we can derive the so-called mass function, 
which is defined as
\begin{equation}
f_c(M) \equiv \frac{M_x^3 \sin^3 i}{(M_x + M_c)^2} = \frac{P_{orb} K_c^3}{2\pi
G},
\label{eq:mf}
\end{equation}
where $M_x$ and $M_c$ are the mass of the compact object and its
companion star respectively, $i$ is the inclination angle of the
binary orbit with respect to the line-of-sight (with $i=0$ for face-on
systems), $P_{orb}$ is the binary orbital period, and $K_c$ is the
semi-amplitude of the optical radial velocity curve. Both $M_x/M_c$ 
and $i$ can be further constrained by modeling ellipsoidal variation 
in the optical light curves~\cite{av78}, and $M_c$ by the spectral 
type of the companion star. For roughly a dozen X-ray binaries, 
$M_x$ is estimated to be greater than $\sim 3 M_{\odot}$, which is 
considered to be a reliable upper limit to the mass of a neutron
star~\cite{rr74}, for any values of $M_c$ and $i$ within their
respective allowed parameter spaces. These sources are, therefore,
good candidates for black hole binaries (BHBs). Table~\ref{tab:bh} 
lists all such systems known to date. It is clear from Eq.~\ref{eq:mf} 
that without making any assumptions the mass function sets a firm 
lower limit to the mass of the compact object. Therefore, those with 
the measured mass function greater than $3 M_{\odot}$ are the best 
BHB candidates (BHBCs).
\begin{table}[btp]
\caption{Stellar-Mass Black Hole Candidates$^{\dag}$}
\vspace{0.4cm}
\begin{center}
\footnotesize
\begin{tabular}{|l|cccc|}
\hline
Sources & $M_x$ & $M_c$ & $i$ & $f_c(M)$ \\ 
 & ($M_{\odot}$) & ($M_{\odot}$) & (degrees) & ($M_{\odot}$) \\ \hline 
Cyg X-1~\cite{her95} & 4.7--14.7 & 11.7--19.2 & 27--67& 0.25 \\ 
LMC X-3~\cite{cow83} & 7--14 & 4--8 & 50--70 & 2.3 \\ 
LMC X-1~\cite{hut87} & 4--8 & 17-24 & 40--63 & 0.14 \\ 
GS 2023+338~\cite{sha94} & 10--15 & 0.5--1.0 & 52--60 & 6.26 \\ 
A 0620-00~\cite{snc94} & 5.1--17.1 & 0.2--0.7 & 31--54 & 3.18 \\ 
GS 1124-68~\cite{snc97} & 4--11 & 0.5--0.8 & 39--74 & 3.1 \\ 
GS 2000+25~\cite{bee96} & 4.8--14.4 & $<$ 0.7 & 43--85 & 4.97 \\ 
GRO J1655-40~\cite{ob97} & 6.8--7.2 & 2.2--2.5 & 69.5 & 3.24 \\ 
H 1705-250~\cite{mar95,har97} & 4.9--7.9 & 0.07--0.42 & 48--51 &4.65 \\ 
GRO J0422+32~\cite{bee97} & $>$ 9 & 0.2--0.4 & 13--31 & 1.21 \\ 
4U 1543-47~\cite{oro98} & 2.7--7.5 & 2.3--2.6 & 24--36 & 0.22 \\ \hline
\multicolumn{5}{l}{$^{\dag}$In case of multiple measurements, the
numbers are taken from the} \\
\multicolumn{5}{l}{most recent papers. See references in these papers 
for proper credits} \\
\multicolumn{5}{l}{on individual results.}
\end{tabular}
\end{center}
\label{tab:bh}
\end{table}

\subsection{Spin of Black Hole}
BHBs are thought to be formed by the core collapse of very massive
stars which are usually fast rotators. The conservation of stellar 
angular momentum would lead to extremely rotating black holes, if 
the progenitor stars behave like rigid bodies. In reality, however, 
the stellar core and envelop are likely decoupled, with the ejected 
envelop carrying away significant amount of angular momentum. There 
may also be other processes that can cause additional loss of angular 
momentum during the formation. Therefore, while we expect that all
black holes rotate to some extent, it is not clear whether any can 
be formed rapidly spinning.

In a binary configuration, the process of mass accretion can, in 
principle, result in the 
accretion of angular momentum by the black hole and, thus, the hole 
can be subsequently spun up~\cite{ba70,th74}. However, most 
BHBs are transient sources (see Table~\ref{tab:bh}, except for the 
top three sources) with a very low duty cycle of X-ray outbursts 
(when the mass accretion rate surges by orders of magnitude), 
so the lifetime-averaged accretion rate is extremely low. Consequently, 
the accretion-induced spin-up of black holes is probably not effective 
in these sources (unlike in active galactic nuclei~\cite{ba70}). Other 
theoretical considerations, such as the presence of sub-Keplerian 
accretion flows in the vicinity of black holes~\cite{ny95,cht95}, would 
only make the mechanism less effective. 

Recently, evidence has been found for the presence of rapidly spinning 
black holes in GRO J1655-40 and GRS 1915+105~\cite{zcc97}, the only
two Galactic sources that occasionally display spectacular radio jets 
with superluminal motion, analogous to radio-loud quasars, which is
why they are sometimes referred to as ``microquasars''. Also
discovered are slowly or moderately spinning black holes in several
``normal'' BHBCs. Not only has the study raised questions 
about the formation of rapidly rotating black holes, it has also
brought into new focus possible
observational consequences of black hole rotation in BHBs. Subsequently, 
a lot of excitement has been generated about the prospect of using
BHBs to probe relativistic effects in strong gravity 
regime~\cite{no97,zcc97,czc98}.

\section{GRO J1655-40: A Test Case}
\subsection{Existence of A Spinning Black Hole?}
\label{ssec:spin}
The binary parameters are known to such a high accuracy for GRO
J1655-40 (with uncertainties typically only a few percent; see 
Table~\ref{tab:bh}) that makes the interpretation of the X-ray data
for this source much more reliable than for any other sources. Despite
being a microquasar, GRO J1655-40 appears as a normal BHBC in terms of
the observed spectral properties. The X-ray spectrum can be well
described by an ultra-soft component at low energies and a power law 
at high energies~\cite{zh97}, which is canonical for
BHBCs~\cite{tl95}. The ultra-soft component is generally attributed to 
the emission from the hot inner region of an accretion disk, while 
the power-law tail is thought to be the product of soft photons being 
Compton upscattered by energetic electrons in the region.

In practice, the ultra-soft component is often modeled by a
multi-color blackbody~\cite{mi84}. In this model, the accretion disk 
is assumed to be geometrically thin, optically thick (i.e., the 
standard $\alpha$ disk~\cite{ss73,nt73}). The emission from such 
a disk is a blackbody, but the temperature of the blackbody is a
function of the distance to the black hole. There are only two free 
parameters in the model: the radial distance of the inner disk edge 
to the black hole and the effective temperature at the inner edge, 
once the distance of the source and the inclination angle of the disk 
(which is usually assumed to be equal to the inclination of the binary 
orbit) are known. The model has worked remarkably well for a number of 
BHBCs (see a review by Tanaka and Lewin~\cite{tl95}). One of the most 
interesting results is that the inferred radius of the inner disk edge 
remains fairly stable and constant as the observed fluxes vary 
greatly~\cite{tl95}, strongly implying that the last (marginally) 
stable orbit is reached.

Since the model assumes Newtonian gravity, appropriate corrections
need to be made to include relativistic
effects~\cite{pt74,ha89}. Also, gravitational shift and 
gravitational focusing cause both the observed color temperature 
and integrated flux to deviate from the local values, depending on 
the inclination angle of the disk and the spin of the black 
hole~\cite{cu75}. Furthermore, in the hot inner region of an 
accretion disk, electron scattering is likely to dominate over 
free-free absorption, so the inner disk actually radiates 
approximately as a ``diluted'' blackbody with the peak effective 
temperature lower than that derived from fitting the
spectrum~\cite{ehs84}. Unfortunately, this ``color correction factor'' 
is still poorly determined, although it appears to depend only weakly 
on the properties of the black hole, mass accretion rate, or the 
location on the disk~\cite{st95}. 

Taking into account these factors, we found that for several BHBCs the
inferred radius of the inner disk edge is consistent with that of the 
last stable orbit around non-rotating black holes~\cite{zcc97}. When 
we applied the same model to the spectrum of GRO J1655-40 during an 
outburst, however, we found that the inner edge of the disk is only
about 1.2 Schwarzschild radii away from the black hole~\cite{zcc97}, 
which is, of course, impossible if the black hole does not rotate. 
Therefore, we were forced to conclude that GRO J1655-40 contains a 
{\it rapidly} spinning black hole. The spin of the black hole was 
determined to be $\sim$93\% maximal rotation for this source.

The ``ordinary'' nature of GRO J1655-40 as a BHBC (in terms of its 
spectral properties) is strongly supported by 
recent observations with the {\it Rossi X-ray Timing Explorer}
(RXTE). The source was monitored by RXTE extensively throughout its 
recent (1996--1997) X-ray outburst. From these observations, Sobczak 
et al.~\cite{sob98} found that the observed X-ray spectrum can be 
well characterized by the canonical spectral shape of BHBCs (soft 
multi-color blackbody plus hard power law) throughout the entire
period, which confirms the previous findings~\cite{zh97} (although 
the actual values of the parameters differ slightly). In addition, 
they found that the inferred radius of the inner disk edge remained 
roughly constant (with occasional low points) as the X-ray flux varied 
by a factor of 3--4, similar to normal BHBCs~\cite{tl95}. 

However, these authors mistakenly concluded that the corrections for
relativistic effects are negligible for this source (actually they
mis-quoted Zhang et al.~\cite{zcc97} on this issue). As a matter of
fact, the general relativistic effects are very significant in this 
case~\cite{zcc97}. In particular, the inner boundary condition (i.e.,
torque free at the last stable orbit) can drastically change the 
temperature profile of the inner portion of the accretion disk under 
Newtonian gravity~\cite{pt74,ha89}. For GRO J1655-40, the
correction amounts to a factor of $\sim$2, after taking into account 
all the effects. Consequently, the radius of the inner disk edge in
Sobczak et al. was over-estimated by the same amount~\cite{sob98}. 

Our interpretation of the same data would, therefore, put the last 
stable orbit roughly at 1.5 Schwarzschild radii, which confirms the 
need for the presence of a rapidly rotating black hole in this system 
(although the spin of the black hole is $\sim$78\% maximal rotation 
in this case). We take note of occasional low values of the inferred 
radius when the measured X-ray luminosity is relatively
high~\cite{sob98}. We interpret them as the times when the adopted 
spectral model breaks down. It is likely that when accretion rate is 
high the inner portion of a standard $\alpha$ disk may become unstable 
to radiation pressure and fragmentation~\cite{le74,kro98}. During the 
outburst, GRO J1655-40 might be situated close to the critical point 
for such instability, as in the case of Cyg X-1 (A. Zdziarski, private
communication). 

In contrast, Sobczak et al. proposed that the last stable orbit was
reached by the accretion disk {\it only} at the {\it smallest} radius 
(i.e., the lowest point) obtained from their spectral analysis. To 
explain why the inner edge of the disk stays far away from the last 
stable orbit during nearly the entire outburst, they invoked strong
radiation drag. Intuitively, however, the radiation drag should be
stronger in the higher luminosity state (when the ``low points'' 
occur), which would be opposite to their line of arguments. Moreover, 
it seems unlikely that the radiation drag is capable of significantly 
affecting gas dynamics in an {\it optically thick} disk (which is 
the most fundamental assumption in the model), since any external
radiation fields (if present) can hardly penetrate the surface of the 
disk and the internal emission percolates through the surface
preferentially in the direction perpendicular to the thin
disk. Finally, from a
practical point of view, it seems risky to base the entire argument on 
a single data point which is barely allowed physically~\cite{sob98}, 
especially considering the fact that not all the ``low points'' reach 
the same level.

In summary, while further studies are required to reveal important 
details about accretion processes in BHBCs, it seems highly plausible, 
as suggested by 
observations~\cite{tl95}, that during an X-ray outburst the accretion 
disk is of the standard form~\cite{ss73,nt73} in transient BHBCs 
and the inner disk edge reaches and remains at the last stable orbit. 
GRO J1655-40 is no exception. Quantitative modeling of its X-ray 
spectrum has led to the conclusion that a rapidly rotating black hole 
is likely to be present in this source. 

\subsection{Gravitomagnetic Precession}
\begin{figure}[t]
\centerline{\epsfig{figure=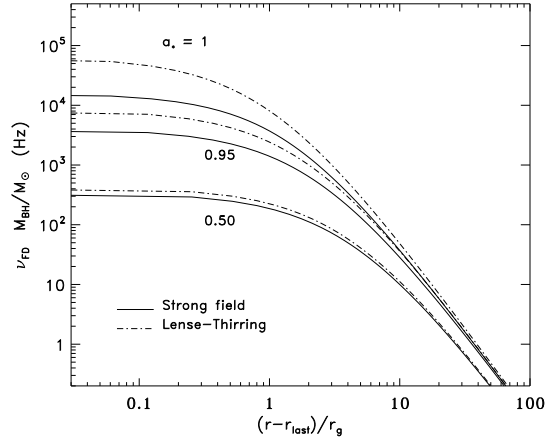,width=3.0in}}
\caption{Gravitomagnetic precession frequency (multiplied by black
hole mass), as a function of the distance from the last stable orbit, 
for different black hole spin. The weak-field limits are also shown 
in dash-dotted line for comparison. }
\label{fig:fd}
\end{figure}

A natural consequence of curved spacetime in the vicinity of a spinning 
black hole is the gravitomagnetic precession of orbits that are tilted 
with respect to the equatorial plane perpendicular to the spin
axis~\cite{lt18}. For circular orbits, which are probably most
relevant to accretion processes in X-ray binaries, the precession
frequency can be expressed as 
$\nu_{FD} = \nu_{\theta} \Delta \Omega/2\pi$, 
where $\nu_{\theta}$ is the orbital frequency of the polar (or $\theta$)
motion, and $\Delta \Omega$ is the angle by which the line of nodes of a
circular orbit are dragged during each orbital period. Note that this
definition is slightly different form the one adopted by Cui et 
al.~\cite{czc98}. As correctly pointed out by Merloni et al.~\cite{mer98}, 
with the new definition, $\nu_{FD}$ approaches the rotation frequency 
of the black hole at the event horizon, which is physically expected. 
Fig.~\ref{fig:fd} shows $\nu_{FD}$ as a function of black hole spin. 
Comparing it with Fig. 1 of Cui et al.~\cite{czc98}, we find that, as 
expected, the difference becomes significant only in the case of 
extremely spinning black holes when the last stable orbit approaches 
the event horizon. Therefore, the results of Cui et al.~\cite{czc98}
hardly change, even for microquasars, neither do any of the
conclusions that they reached (see also discussion in the following 
section). 

\subsection{``Stable'' QPOs}
Quasi-periodic oscillations (QPOs) have been observed in the X-ray light
curves of BHBCs over a wide frequency range, from mHz to roughly a few 
hundred Hz (reviews by van der Klis~\cite{van95} and Cui~\cite{cui98}). 
While it is clear that no single process can account for all QPOs, the 
origins of QPOs are not known. 

One type of QPOs in BHBCs is of particular interest here. Such QPO was 
first reliably established in GRS 1915+105 with the detection of a 
67 Hz QPO~\cite{mrg97}. One unique characteristic of this QPO 
is the stability of its frequency against any variation in the X-ray flux. 
The QPO appears to represent a transient event, and seems to be
present only in certain spectral states. Shortly after this discovery, 
a similar 
signal was detected at $\sim$300 Hz for GRO J1655-40~\cite{rem98},
{\it only} when the X-ray luminosity was relatively high~\cite{sob98}
(i.e., at the ``low points'', as discussed in \S~\ref{ssec:spin}). 
We subsequently proposed that these ``stable'' QPOs are perhaps the
observational manifestation of gravitomagnetic precession of accreted,
orbiting matter around spinning black holes~\cite{czc98}.

\begin{figure}[t]
\centerline{\epsfig{figure=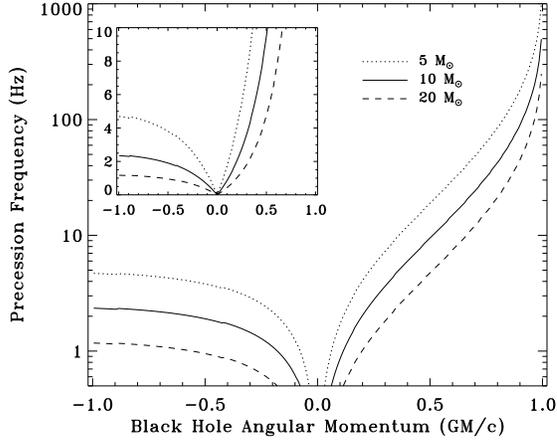,width=3.0in}}
\caption{Gravitomagnetic precession frequency of a disk annulus, at
which the effective disk temperature peaks, as a function of black 
hole spin (assuming $Q=1$). }
\label{fig:fdspin}
\end{figure}

To facilitate comparison with observations, we have computed the
expected gravitomagnetic precession frequency of a disk annulus at
which the effective disk temperature peaks. The results are plotted in
Fig.~\ref{fig:fdspin}, as a function of black hole spin for three
values of black hole mass.
Comparing this figure to Fig. 2 of Cui et al.~\cite{czc98}, we find,
again as expected, significant difference exists only for large (and
positive) black hole spin. Note that the results depend little on the
value of Q~\cite{czc98,mer98}, which is a constant of motion 
that specifies the tilt of precessing orbits~\cite{wil72}.

For GRO J1655-40 (where $M_{bh} = 7 M_{\odot}$), the spin of the black 
hole would be about 97\% maximal rotation were the observed 300
Hz QPO to be attributed to the gravitomagnetic precession (as shown 
in Fig.~\ref{fig:fdspin}). This result assumes that the QPO originates 
in the modulation of the X-ray emission from the disk, which does not
necessarily have to be the case, as will be discussed in
\S~\ref{ssec:xmm}. To
derive a lower limit to the black hole spin, we computed the frequency
of gravitomagnetic precession right at the last stable orbit, as a
function of the spin, and set the result equal to the QPO
frequency. We found that in this case the spin would be $\sim$87\% 
maximal rotation. The results are consistent with those obtained 
spectroscopically (see also Fig.~\ref{fig:ms}). 

\section{Application to Other Sources}
\label{sec:aos}
Attempts have been made to apply the model to QPOs observed in ``normal''
BHBCs~\cite{czc98}. The closest examples are perhaps 
the so-called ``very-high-state'' QPOs that were observed only twice,  
in GX 339-4~\cite{miy91} ($\sim$6 Hz) and GS 1124-68~\cite{bel97} 
(5--8 Hz), in a high luminosity state (i.e., the very-high state).
The frequency of such QPOs is fairly stable against flux variation, 
and any change in the QPO frequency can be explain by the variation 
in radiation pressure~\cite{czc98}. The low values of the QPO
frequency would imply the presence of only slowly rotating black holes 
in these sources were these QPOs to be produced by the gravitomagnetic 
precession of accreted matter. This would in turn support the
speculation that the relativistic jets observed in microquasars might
somehow related to the spin of black holes~\cite{zcc97}.

Similar QPOs have also been observed in GS 1124-68~\cite{bel97} and Cyg
X-1~\cite{cui97} during the transition between the low and high state.
However, in these cases, the situation is much more complicated and,
thus, the interpretation is much more uncertain, because of the
variable nature of the sources and the transient nature of the QPO
phenomenon in general~\cite{cui98}. For a particular source, a 
different set of QPOs are often present at different times (or
fluxes). It is, therefore, difficult to associate one QPO (if any)
with the gravitomagnetic precession of accreted matter. For instance, 
a lower-frequency QPO (6--8 Hz) was often present in GRO J1655-40 
during the same outburst when the 300 Hz QPO was
discovered~\cite{men98}. To make further progress, an systematic 
effort is imperative to reliably characterize and classify QPOs 
in BHBCs. 

\section{Unsettled Issues}
\label{sec:ui}
\subsection{Origin of Tilted Orbits} 
Accreted matter must be in a non-equatorial orbit, with respect to 
the spin axis of the black hole, in order for the orbit to undergo 
gravitomagnetic precession. In principle, such orbits could be 
produced during the formation of a black hole binary, due to the 
misalignment of the spin axis of the black hole with the axis of the 
binary motion~\cite{hil83}. 

If mass accretion proceeds through a flat viscous disk, the disk can
still be warped, due to irradiation by a central X-ray 
source~\cite{pri96}. However, in this case, differential 
gravitomagnetic precession generates twists in the disk and the inner 
region of the disk tends to gradually realign with the equatorial 
plane due to the viscous torque~\cite{bp75}. Once the realignment
takes place, gravitomagnetic precession modes are damped out and,
thus, the QPO should simply disappear. On the other hand, this process 
can help explain why the QPOs seem to be present {\it only} during 
transitional or unstable periods, when the inner region of the disk 
experiences significant changes.

As mentioned in \S~\ref{ssec:spin}, the inner portion of the disk may
become unstable~\cite{le74,kro98}, when the accretion rate is
relatively high. Once this occurs, the disk might become fragmented. 
The gravitomagnetic precession of the fragments is probably only 
weakly damped, so it might be capable of producing the observed QPOs.

\subsection{Excitation of Gravitomagnetic Modes}
A recent theoretical development involves the discovery of a class of
high-frequency gravitomagnetic modes that are only weakly damped, in 
addition to the strongly-damped low-frequency ones~\cite{ml98}. These
high-frequency modes are very localized spiral corrugations of the
inner portion of the disk. They are perhaps most relevant to the
observed QPOs, because X-ray emission from BHBCs is
likely to originate very close to the central black hole. However, it
remains to be seen whether the modes can be excited easily and whether
they are capable of modulating X-ray emission~\cite{ml98}.

It should be pointed out that two critical assumptions are made in the
study: small tilt angle and (pseudo) Newtonian potential. The former 
simplifies the equation by ignoring the non-linear aspect of the
problem, which might affect the damping time scales of gravitomagnetic 
modes in a significant way; non-linear calculations are required to 
shed light on this issue. The latter might still represent a good 
approximation for systems that contain a relatively slowly rotating 
neutron star, but is certainly invalid for rapidly rotating black
holes in which the stable QPOs are observed. These assumptions,
therefore, tend to limit the scope of the applicability of the
results. At present, it would seem premature to draw any definitive 
conclusions about confirming or rejecting the gravitomagnetic origin 
of the stable QPOs in microquasars.

\subsection{X-ray Modulation Mechanisms}
\label{ssec:xmm} 
The gravitomagnetic precession of accreted matter only provides a 
natural frequency for the QPOs. To actually see them, certain physical 
processes are required to produce X-ray modulations. The processes 
are entirely unknown at present. Possibilities include (by no means
exclusively): (1) variation in the X-ray emitting area of the 
disk, due to precession; (2) Doppler or gravitational shift of photon 
energy, due to the Keplerian motion of self-illuminating clumps or 
``hot spots'' in the accretion disk; (3) oscillation that modulates 
the emission from the disk; and (4) occultation of a highly compact 
X-ray emitting region by disk fragments or the inner part of the 
disk itself.

The transient nature of the QPOs seems to imply that the
first possibility, if viable at all, must be only a small effect. Any
processes that involve discrete clumps would unavoidably also produce 
X-ray modulation at the orbital (both polar and azimuthal)
frequencies~\cite{ms98,mer98}. These QPOs have not been seen 
in BHBCs, despite the high quality of RXTE data (the QPOs of comparable 
strength have been detected at around kHz, using the RXTE data, in 
more than a dozen neutron star systems~\cite{van97}). Therefore, such 
processes also seem unlikely to be
responsible for the observed QPOs. Hot spots or oscillations that
modulate the emission from the accretion disk can also modulate
Comptonized hard X-ray emission, if the seed photons originate in 
the disk emission. However, it seems difficult for these processes 
to amplify the fractional amplitude of the QPOs toward high photon 
energies, which is observed~\cite{mrg97,rem98}. Such energy
dependence of the QPO amplitude might, in our view, hold the key to
our understanding of the origin of the QPOs.

The only remaining possibility in the above list is the occultation of
a highly compact hard X-ray emitting region by detached disk annuli 
or the inner disk itself. Interestingly, both
observations~\cite{gro98} and theories~\cite{na97,lt98} support the 
presence of such X-ray emitting 
regions in transient BHBCs during an X-ray outburst. As an
example, Laurent \& Titarchuk~\cite{lt98} recently showed that the 
observed power-law type hard X-ray spectrum of BHBCs may be the result 
of soft photons from the accretion disk being inverse-Compton
scattered by the bulk motion of relativistic electrons in the vicinity
of central black holes. The Comptonizing electrons are spatially confined 
almost entirely within the inner edge of the disk. In the context of 
this model, harder X-rays should, on average, originate closer to the 
black hole. Therefore, a precessing ring of matter occults a larger 
fraction of the harder X-ray emitting region, which might explain the 
observed energy dependence of the QPO amplitude, although a definitive 
answer still awaits detailed calculations based on this model.

\subsection{Alternative Models} 
The stable QPOs in GRS 1915+105 and GRO J1655-40 can conceivably be 
associated with the Keplarian motion of self-illuminating clumps 
right at the last stable orbit around the black 
hole~\cite{mrg97,rem98}. In the case of GRO J1655-40 whose
black hole mass is accurately known, the model would be able to
explain the observed QPO frequency (300 Hz) only if the black hole 
does not rotate~\cite{rem98}. This is clearly incompatible with the
requirement for the presence of a rapidly rotating black hole in this
source from modeling the observed X-ray spectrum. A more serious (and
model independent) problem with this interpretation is due to the 
fact that general relativistic effects cause the temperature of the 
accretion disk to vanish at the last stable
orbit~\cite{pt74,ha89}. Consequently, no X-ray emission is expected 
from the inner edge of the disk.

Another proposal has been made to associate the stable QPOs to 
epicyclic oscillation modes in the accretion disk~\cite{now97}. Not 
only can such modes be supported by the disk, they also become trapped 
in the innermost portion of the disk, purely due to relativistic
effects~\cite{kf80,nw92,nw93,ct95,per97}. It is also 
realized that g-mode oscillations affect the largest area of a region 
in the disk where most X-rays are emitted and are, therefore, perhaps 
most relevant to observations~\cite{now97}. This model can explain 
the measured frequencies of the stable QPOs in microquasars. However,
like other disk-based processes, it may have difficulty in explaining
the observed energy dependence of the QPO amplitude. Furthermore, it  
cannot account for the very-high-state QPOs or other relatively 
low-frequency QPOs in BHBCs (see \S~\ref{sec:aos}), although arguments 
can be made that these are fundamentally different from the stable 
QPOs in microquasars.

\begin{figure}[th]
\centerline{\epsfig{figure=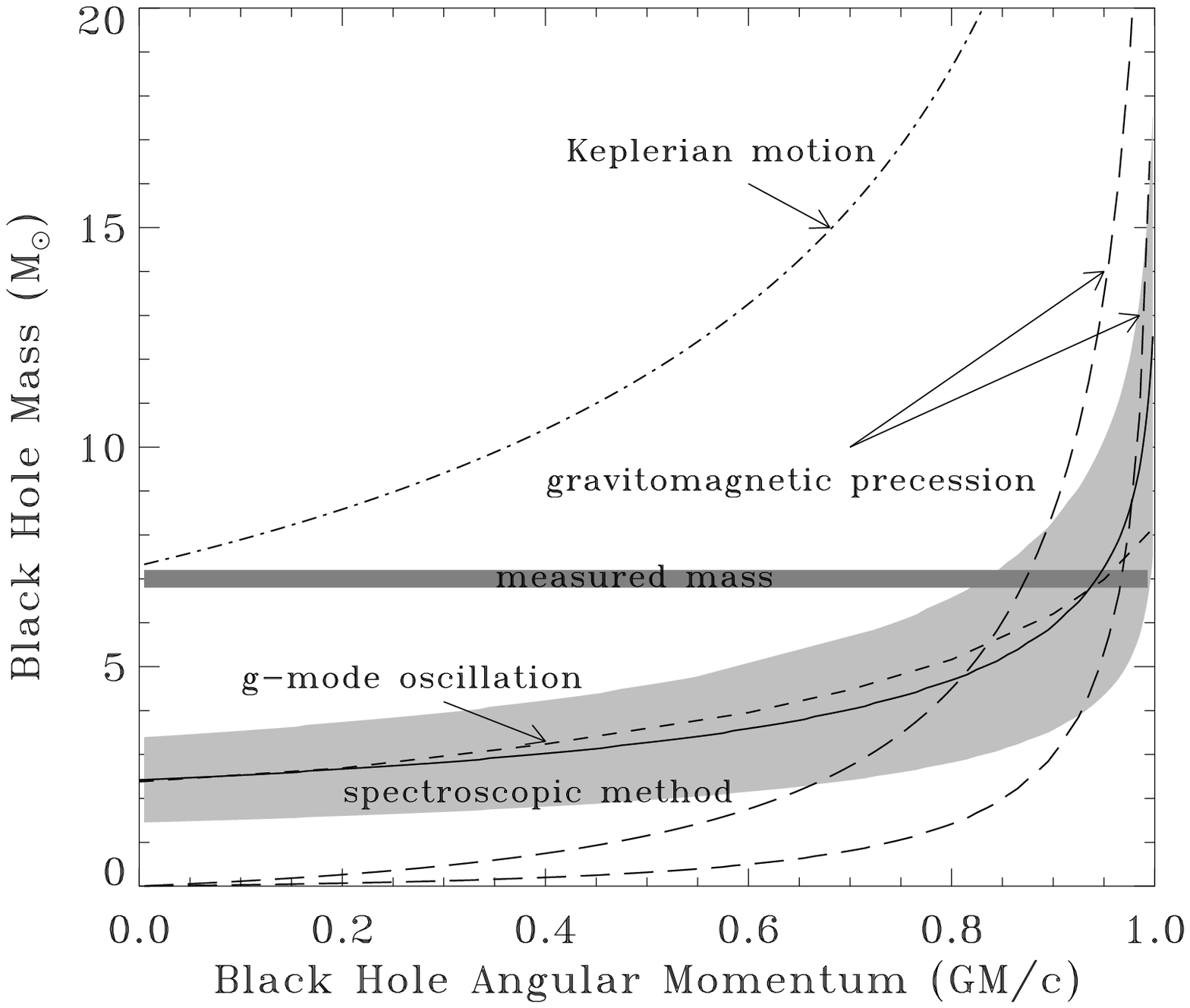,width=2.8in}}
\centerline{\epsfig{figure=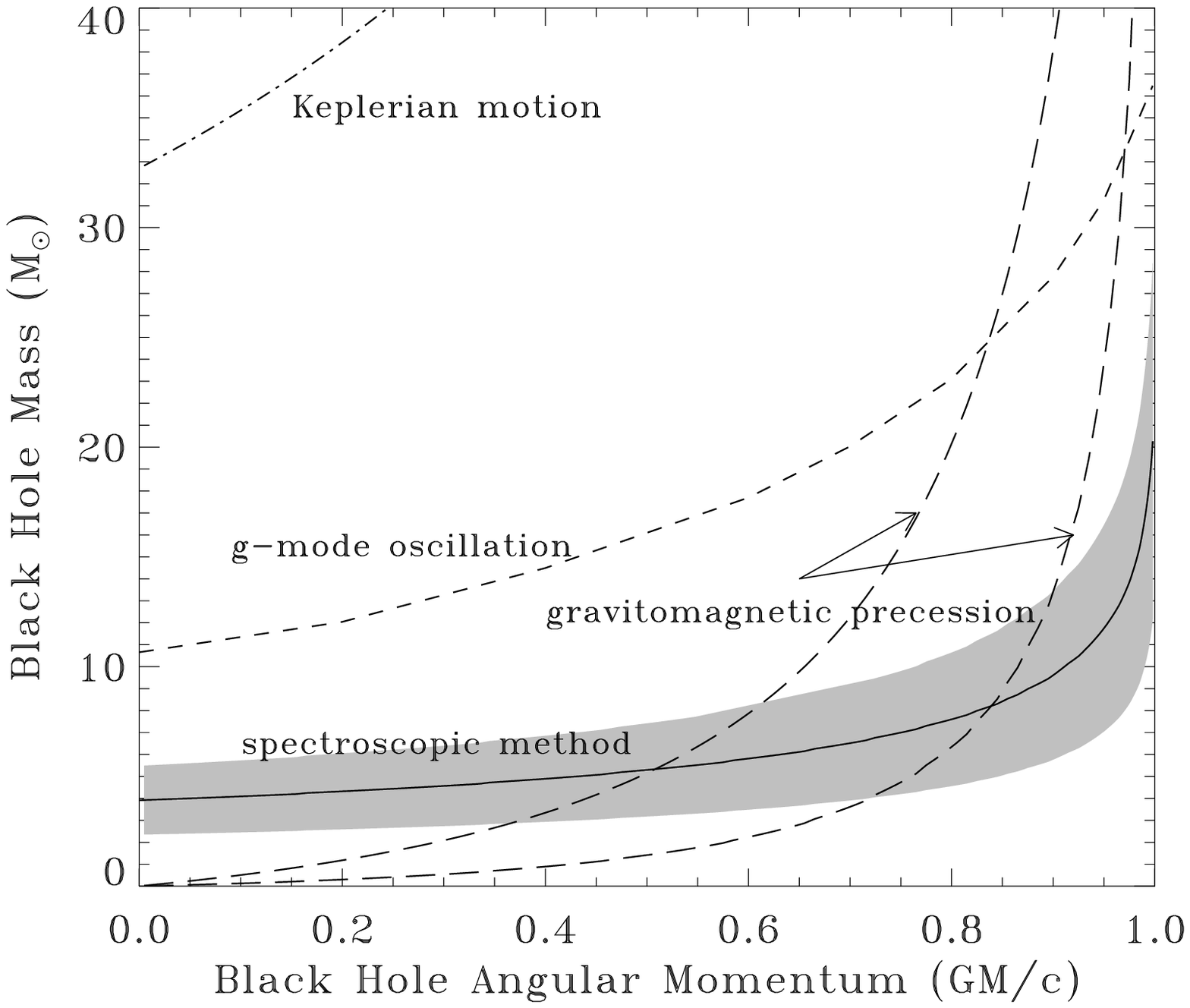,width=2.8in}}
\caption{(top) Allowed parameter space for the mass and spin of the 
black hole in GRO J1655-40. The solid line shows the spectroscopic
results of Zhang et al. (see \S~2.1), with the 
lightly-shaded region indicating the estimated uncertainty. The 
dot-dashed, short-dashed, and long-dashed lines show, respectively, the 
predictions of three proposed models: Keplerian motion, g-mode
oscillation, and gravitomagnetic precession. For the last model,
two limiting cases are shown for frequencies of gravitomagnetic 
precession at the inner edge of the disk (upper curve) and at where 
the effective disk temperature peaks (lower curve). The heavily-shaded 
area represents the confidence region of the measured black hole 
mass. (bottom) Similar to the top panel, but for GRS 1915+105. In 
this case, the mass of the black hole is not known. }
\label{fig:ms}
\end{figure}

To quantify the comparison among the three possible models
(gravitomagnetic precession, Keplerian motion, and g-mode
oscillation), we have applied the models to both GRO J1655-40 and GRS
1915+105. For each model, a relationship between the mass and spin of
the black hole is derived from the measured frequency of the QPO. The
results are shown in Fig.~\ref{fig:ms}, for both sources. In the
gravitomagnetic precession model, two limiting cases are shown for 
frequencies at the inner edge of the disk and at where the effective
disk temperature peaks, since we do not know the actual X-ray
modulation mechanism (see \S~\ref{ssec:xmm}). Also shown in
the figure is the same relationship derived from modeling the observed
X-ray spectrum for each source~\cite{zcc97}. The intersection between 
the latter
and each of the three model curves, therefore, yields a solution to
the mass and spin of the black hole, for a particular source, as 
predicted by that model. For GRO J1655-40, the mass of the black hole
is known quite accurately, so it can be used to determine which models
are compatible with the observed properties.

The uncertainty of the spectroscopic results
is likely dominated by that of the color correction factor (up to
$\sim$12\%~\cite{st95}), which can contribute as much a uncertainty
as $\sim$24\% in the derived radius of the last stable
orbit~\cite{zcc97}. The next major source of uncertainty comes from 
determining the source distance, to which the radius of the last stable
orbit is directly proportional. It amounts to $\sim$6\% for 
GRO J1655-40~\cite{hr95} and $\sim$12\% for GRS 1915+105~\cite{mr94}. 
Likely minor contributions include uncertainties in the binary 
inclination angle, in the derived X-ray flux and the peak effective 
temperature of the disk, and in various relativistic correction 
factors~\cite{zcc97}. To be relatively conservative, we have adopted 
an overall uncertainty of 40\% on the radius of the last stable orbit
for both sources. The lightly-shaded area in Fig.~\ref{fig:ms}
reflects this measurement uncertainty.

Clearly, the Keplerian motion model is ruled out, since it is far 
from being consistent with the spectroscopic results. While the 
g-mode oscillation model seems to work for GRO J1655-40, it fails to 
yield any physical solutions for GRS 1915+105. The difference is so
large that any reconciliation between the model and the data seems
unlikely, unless the spectroscopic method has severely underestimated 
the radius of the last stable orbit (or the method does not apply in 
this case). On the other hand, the gravitomagnetic precession model
seems to explain the observations well for both sources. For GRS
1915+105, the model constrains the mass ($\sim$3--14 $M_{\odot}$) and 
spin ($\sim$37--90\% maximal rotation) of the black hole. Future
infrared observations of the source might be able to test the mass
constraint. 

\section*{Acknowledgments} We would like to thank Andrea Merloni for
sharing with us the results from his work prior to publication. W. Cui  
is also grateful to Ron Remillard for keeping him updated of the
results on GRO J1655-40 and for many helpful discussions, despite 
some difference in their interpretation of the data. This work is 
supported in part by NASA through Contract NAS5-30612.

\end{document}